\documentclass[a4paper,11pt]{article}
\usepackage{pos}

\title{Monitoring Gamma-Ray Burst VHE emission with the Southern Wide-field-of-view Gamma-ray Observatory}
 \ShortTitle{GRB monitoring with SWGO}

\author*[a]{G. La Mura}
\forColl{SWGO}

\affiliation[a]{Laborat\'orio de Instrumenta\c{c}\~ao e F\'{i}sica Experimental de Part\'{i}culas, Av. Prof. Gama Pinto 2, Lisboa, Portugal}

\emailAdd{glamura@lip.pt}

\abstract{It has been established that Gamma-Ray Bursts (GRB) can produce Very High Energy radiation (E > 100 GeV), opening a new window on the investigation of particle acceleration and radiation properties in the most energetic domain. We expect that next-generation instruments, such as the Cherenkov Telescope Array (CTA), will mark a huge improvement in their observation. However, constraints on the target visibility and the limited duty cycle of Imaging Atmospheric Cherenkov Telescopes (IACT) reduce their ability to react promptly to transient events and to characterise their general properties. Here we show that an instrument based on the Extensive Air Shower (EAS) array concept, proposed by the Southern Wide Field-of-view Gamma-ray Observatory (SWGO) Collaboration, has promising possibilities to detect and track VHE emission from GRBs. Observations made by the Fermi Large Area Telescope (Fermi-LAT) identified some events with a distinct spectral component, extending above $1\,$GeV or even $10\,$GeV, which can represent a substantial fraction of the emitted energy and also arise in early stages of the process. Using models based on these properties, we estimate the possibilities that a wide field of view and large effective area ground-based monitoring facility has to probe VHE emission from GRBs. We show that the ability to monitor VHE transients with a nearly continuous scanning of the sky grants an opportunity to access simultaneous electromagnetic counterparts to Multi-Messenger triggers up to cosmological scales, in a way that is not available to IACTs.}

\FullConference{37$^{\rm{th}}$ International Cosmic Ray Conference (ICRC 2021)\\
		July 12th -- 23rd, 2021\\
		Online -- Berlin, Germany}

\newcommand{\de}{\mathrm{d}}

\begin{document}
\maketitle

\section{Introduction}
Since the time of their discovery, Gamma-ray Bursts (GRBs) have always represented a major astrophysical challenge, both in terms of observations and of theoretical interpretation. After several decades of investigation, it is well established that GRBs are produced within ultra-relativistic jets of material accelerated by the rapid accretion process that occurs during the formation of a {\it magnetar} or a black hole (BH) \cite{Woosley93}. These objects can either result from the core collapse of a very massive star ($M \geq 20\, \mathrm{M}_\odot$), or as the consequence of a binary neutron star (NS) or a NS-BH merger. In spite of sharing a common engine, the two different scenarios are characterized by distinct properties.

Generally speaking, a GRB exhibits two emission regimes: a prompt stage, characterized by variable pulse-like emission, and an afterglow, featuring a smooth time evolution, usually in the form of a power-law or a broken power-law. The distribution of prompt emission duration is bi-modal and it well reflects the existence of two possible origins \cite{Mazets81, Norris84}. The events with prompt emission faster than $2\,$s are named {\it short} GRBs and are considered the counterpart of compact binary mergers, while those lasting from several seconds to minutes are called {\it long} ones and are typically associated with stellar collapse mechanisms \cite{Kouveliotou93}. The afterglow, on the other hand, can be emitted for much longer time-scales, reaching up to hours or even days, and it can be more easily associated with possible counterparts at different frequencies (most commonly including X-rays, optical and radio emission).

The energetic output of GRBs spans between $10^{50}\,$erg and $10^{54}\,$erg, with luminosities as high as $10^{52}\, \mathrm{erg\, s^{-1}}$. A large fraction of the energy is emitted in the spectral range running from hundreds of keV to hundreds of MeV and it usually comes in the form of a combination of smoothly connected power-laws, represented by the so-called {\it Band function} \cite{Band93}. Thanks to the data collected by the {\it Fermi} Large Area Telescope ({\it Fermi}-LAT) \cite{Atwood09}, it has been demonstrated that GRBs can additionally produce photons at energies well above $10\,$GeV and that they should be intrinsically able to radiate above $100\,$GeV, in the Very High Energy (VHE) regime \cite{Ajello19}, as subsequently confirmed by observations carried out with the MAGIC and H.E.S.S. telescopes \cite{MAGIC19, Abdalla19}. It is expected that observations with the Cherenkov Telescope Array (CTA) \cite{CTA19} will further clarify the VHE properties of GRBs, but, due to the short duration of the prompt emission, any follow-up observation will be more likely probing the afterglow phase, rather than the prompt one, missing fundamental information on the VHE emission at the earliest stages.

Here we discuss the potential that a VHE monitoring facility based on the Extensive Air Shower (EAS) detector array concept, such as the Southern Wide Field-of-view Gamma-ray Observatory (SWGO) \cite{SWGOpaper}, has to contribute in the detection of the most energetic emission of GRBs. We organize our discussion as follows: in \S2 we provide an overview of the observational properties of the VHE emission of GRBs and of their theoretical implications; in \S3 we consider the potential contributions of SWGO; finally, in \S4 we summarize our conclusions.

\section{VHE emission from Gamma-Ray Bursts}
There is overwhelming evidence that GRBs are produced as the result of relativistic shocks in a highly magnetized environment. The presence of charged particles and the observed energies imply that synchrotron radiation and inverse Compton (IC) scattering processes are unavoidable ingredients of the expected emission. While relatively simple arguments, concerning the intrinsic opacity to the highest energy photons and their arrival time distribution, tend to agree with the idea that GRB sources can achieve very high bulk Lorentz factors ($\Gamma \geq 300$) and that the most energetic photons appear to be the result of reprocessed emission, no simple theoretical interpretation is able to explain the spectral and temporal complexity of GRBs. The large degree of variability of the prompt stage, seen down to millisecond time-scales, and the particle energy distributions, needed to justify the observed spectra, pose challenging problems, such as those connected with the electron cooling time (e.g. \cite{Preece02}). It has been suggested that other exotic radiation mechanisms could be present. Models including hadronic photo-production \cite{Vietri97} and proton synchrotron emission \cite{Ghisellini20}, for example, appear to provide better interpretations, at least for some more energetic events. If this is the case, characteristic VHE signatures are expected in the spectra \cite{Piran04}.

\begin{figure}
    \centering
    \includegraphics[width=0.6\textwidth]{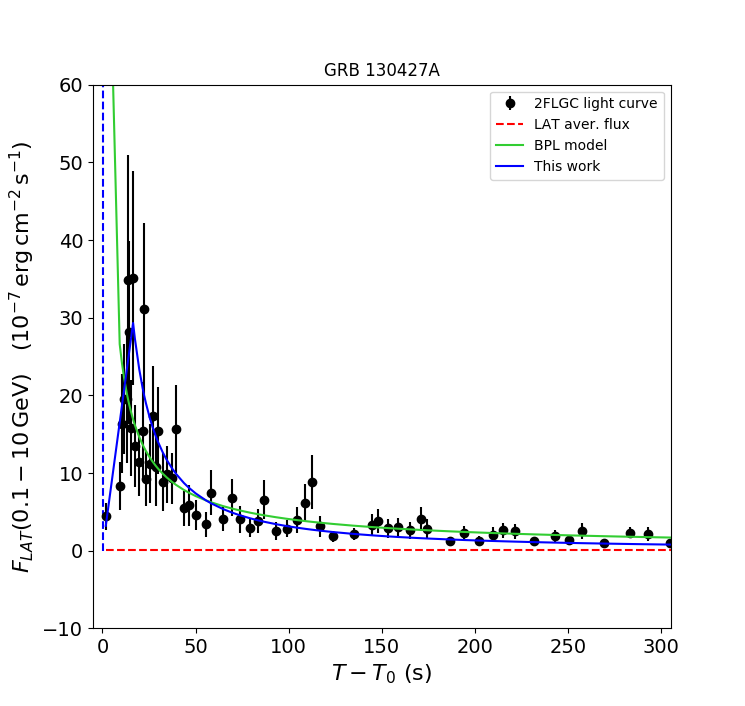}
    \caption{High energy light curve of the GRB~130427A, as observed by {\it Fermi}-LAT, during the first $5\,$min in the $0.1-10\,$GeV energy band. The points with error bars are LAT flux measurements, the green continuous line is the broken power-law fit of the GRB in 2FLGC, the blue continuous curve is the flux adopted in our work, while the vertical blue dashed line is the start time of LAT analysis. The red horizontal dashed line is the average energy flux measured by LAT over the entire duration of the burst. \label{figLC}}
\end{figure}
During the first $10$ years of its monitoring campaign, {\it Fermi}-LAT detected photons in the spectral range $0.1-10\,$GeV for a total of $184$ GRBs, listed in the Second Fermi-LAT GRB Catalogue (2FLGC) \cite{Ajello19}, occasionally recording high energy photons above $10\,$GeV. However, the rather small LAT collecting area ($\sim 1\, \mathrm{m^2}$), combined with the limited duration of the GRB signal, does not allow the instrument to effectively constrain the VHE domain. In the case of the brightest events, like for example GRB 130427A, illustrated in Fig.~\ref{figLC}, LAT tracked the existence of an additional radiation component, with respect to the low-energy Band emission, whose spectral and temporal properties could be well represented in the form of power-law or broken power-law functions, without any evidence of high energy cut-offs. On the contrary, when simultaneous LAT and VHE observations were possible, evidence for an energetic emission in excess of the high energy extrapolation of the LAT spectra was found \cite{MAGIC19, Acciari21}. Follow-up observations, carried out by MAGIC and H.E.S.S., were able to detect VHE emission in the afterglow of some long GRBs, with hints of a VHE signal also from the short GRB~160821B \cite{Acciari21}. The presence of VHE emission in both event classes is difficult to explain in one-zone models of Synchrotron-Self-Compton (SSC) emission from external shocks and it favours alternative scenarios where TeV scale emission can be relatively enhanced. This suggests that monitoring instruments with large collecting areas may provide fundamental insight in the VHE properties of GRBs, down to the time-scale of the early afterglow and of the prompt emission. In the following, we explore the scientific opportunities expected from a comparison of GRB observations, and of their possible extension to the VHE domain, with the sensitivity requirements for a survey instrument, like the one proposed by the SWGO Collaboration.

\section{GRB monitoring with SWGO}
Due to the typically very low photon fluxes, the observation of VHE sources emitting for short times requires the use of large instrumented areas. These can only be attained by ground-based facilities, where IACT observatories achieve the best performance, thanks to their good angular resolution and to their excellent background rejection power. However, the narrow Field of View (FoV), of the order of few square degrees, and the limitation to operate during clear nights, pose serious constraints on the possibility that Cherenkov telescopes may quickly and effectively cover fast transients, such as the prompt emission of GRBs. Observations with EAS arrays, on the contrary, grant a continuous scanning of a much larger FoV ($\sim 1\,$sr), although they need to cope with a larger rate of cosmic-ray background events and they have to be located at high altitude ($\geq 4500\,$m a.s.l.), in order to be able to detect particles from showers produced by primaries in the sub-TeV energy domain.

\begin{figure}
    \centering
    \includegraphics[width=0.9\textwidth]{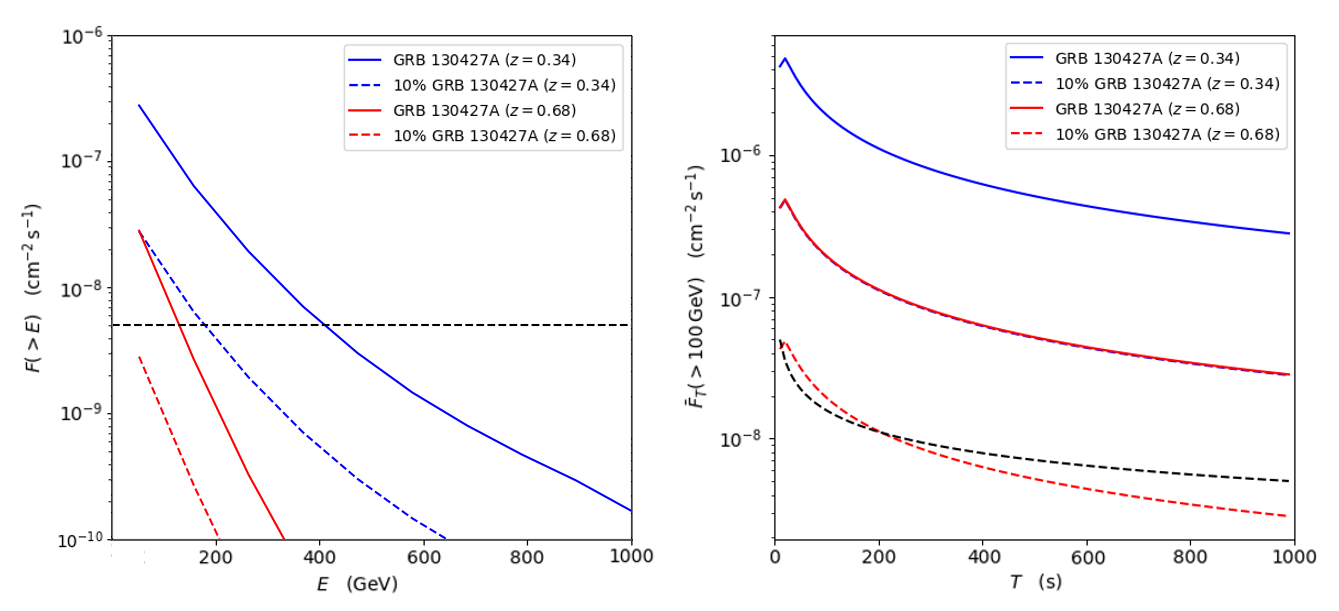}
    \caption{Left panel: average photon flux expected above any given energy in $1000\,$s for a burst with the spectral and temporal characteristics of GRB~130427A (blue continuous line), an identical burst with $10\%$ its strength (blue dashed line) and the same previous cases computed for twice the measured redshift (red continuous and dashed lines). The horizontal black dashed line represents a reference flux limit of $5 \cdot 10^{-9}\, \mathrm{ph\, cm^{-2}\, s^{-1}}$. Right panel: same as left panel, but calculated for the instantaneous integral photon fluxes, expected above $100\,$GeV and assuming that the limiting flux for a detection scales approximately as the square root of the observation time. \label{figInstrument}}
\end{figure}

The high energy emission observed from GRBs, above the GeV scale, can usually be expressed in the form of a power-law spectrum, such as:
\begin{equation}
    \dfrac{\de N(t)}{\de E} = N_0(t) \left( \dfrac{E}{E_0} \right)^{-\alpha} \exp [-\tau(E, z)], \label{eqSpec}
\end{equation}
where $N_0(t)$ is the photon flux per unit energy at time $t$ and energy $E_0$ $\alpha$ is the spectral index (generally found in the range $1.5 \leq \alpha \leq 3$ with an average value close to $2$) \cite{Ajello19}, and $\tau(E, z)$ is the pair production opacity that high energy $\gamma$-rays experience mainly on the Extra-galactic Background Light (EBL) photons. 
The expected VHE $\gamma$-ray fluxes can be estimated at different times from Eq.~(\ref{eqSpec}), in the assumption that the spectrum normalization scales with time, according to a broken power-law function, defined as:
\begin{equation}
    N_0(t) = N_{p} \left( \dfrac{t - T_0}{T_{p} - T_0} \right) \: \mathrm{for}\: T_0 \leq t \leq T_{p} \quad \mathrm{and} \quad
    N_0(t) = N_{p} \left( \dfrac{t}{T_{p}} \right)^{-\delta} \: \mathrm{for}\: t > T_{p}, \label{eqLC}
\end{equation}
where $N_p$ is the peak flux normalization, $T_0$ is the starting time of the GRB emission, $T_p$ the time required to achieve peak luminosity, and $\delta$ is the index of the subsequent power-law decay. The amount of expected photon fluxes can be obtained by integrating the spectra of GRBs either in time or in energy, under the assumption of a specific EBL opacity model \cite{Dominguez11}. As it is shown by Fig.~\ref{figInstrument}, the resulting fluxes are controlled by the burst own luminosity, by the redshift of the source, and by the energy window considered in the observation. The data collected by {\it Fermi}-LAT and the subsequent detection of VHE components in the spectra of some events by MAGIC and H.E.S.S. show that such high energy spectral components can appear both in long and short GRB, suggesting that the emission may be interpreted in terms of internal radiation mechanisms, rather than through environmental interactions. The LAT observations, in particular, form a good basis to model the temporal evolution of the energetic component, although information on the prompt and early afterglow emission is still very scarce, being only available for a few particularly bright events.

Using the spectral and the light curve information provided by 2FLGC, we can estimate the photon fluxes, detected by {\it Fermi}-LAT in the $0.1-10\,$GeV energy range as a function of time, and attempt a calculation of the VHE extrapolation. If the redshifts of the observed GRBs were known, this operation would be possible, with the only uncertainty related to the choice of the adopted EBL opacity model. In practice, however, only $34$ LAT detected GRBs, listed in 2FLGC, have an actual measurement of the redshift, so, in order to take into account the effects of EBL opacity, when moving from the LAT observed window to the VHE domain, we need to test the visibility of sources with different intrinsic power, spectral index and redshift. A detailed study of the possible effects of redshift distribution is currently in progress. Here we limit ourselves to consider what kind of GRBs would be detectable, if the spectral window covered by a ground based monitoring instrument can be extended in the sub-TeV energy range down to approximately $100\,$GeV.

The situation illustrated in Fig.~\ref{figInstrument} shows that an integrated flux sensitivity of $5 \cdot 10^{-9} \mathrm{ph\, cm^{-2}\, s^{-1}}$ for approximately $1000\,$s observing time, pursued by SWGO, would result in excellent detection possibilities for the characteristics of the highest energy LAT detected GRB. More realistic expectations, however, should not only limit to consider the best available case. For this reason, the comparison of expected fluxes and desired sensitivity is carried out in a few additional cases, where the effects of lower intrinsic power or higher source redshift are also taken into account. As it is clearly illustrated by the diagrams, the largest amount of flux is expected close to the low energy threshold. The indicated flux sensitivity limit offers very good prospects for the detection of VHE emission from bright GRBs or fainter events at moderate redshift values ($z < 0.5$). Intrinsically luminous events can be effectively tracked at higher redshift ($z \geq 0.5$), while the observation of fainter GRBs would be obviously more problematic, although we can still appreciate the possibility of a marginal detection of a relatively low-power, high redshift event, provided that the peak of its light-curve occurs when the source is visible. Using a wide FoV monitoring instrument, therefore, we would have the completely new opportunity to explore the radiation mechanisms connected with the prompt emission of a population of targets not limited to only the brightest events.

Taking into account the broad range of radiated power, spectral forms, event duration and redshift distribution, the explored parameter space roughly corresponds with the possibility to detect $10\%$ of the brightest {\it Fermi}-LAT observed GRBs, which would averagely result in one likely VHE GRB detection per year. In addition to represent a substantial improvement over the results achieved in the past years, this level of desired performance would grant the opportunity to detect the transient in very short times (few seconds or less), providing invaluable simultaneous coverage to Multi-Messenger triggers and acting as an excellent complementary facility to the follow-up investigations planned for CTA.

\section{Conclusions}
The systematic study of GRBs in the VHE domain will represent a fundamental step in the quickly evolving field of Multi-Wavelength and Multi-Messenger Astrophysics. If the first detections of Gravitational Waves (GW) marked a corner stone in scientific research \cite{Abbott16}, the associated observation of GW~170817 and of the short GRB~170817A represented the first direct evidence of the compact binary merger as a viable explanation to short GRBs, paving the way to a wealth of Cosmological and High Energy Physical tests \cite{Abbott17a, Abbott17b}. A fundamental question related to the nature of GRBs and to their distinction in classes is whether the formation of the jet and its subsequent evolution involve substantial hadronic processes, a matter that could be unambiguously solved by the association of GRBs with neutrino events or with early VHE emission (see e.g. \cite{Bustamante17}). 

\begin{figure}
    \centering
    \includegraphics[width=0.9\textwidth]{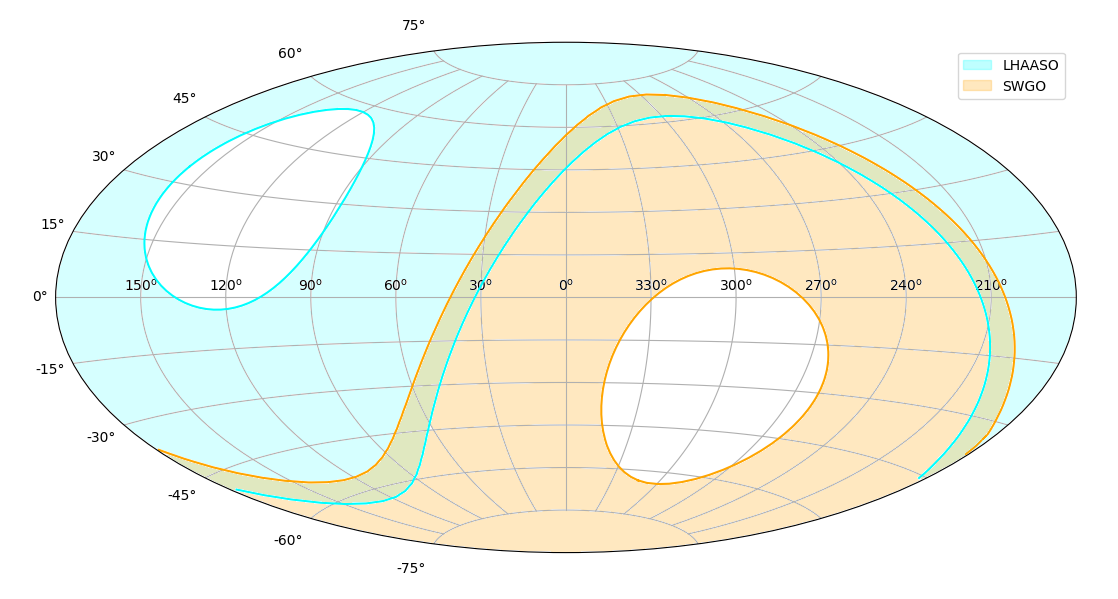}
    \caption{Representation of the visible sky, within $30^{\rm o}$ from zenith, showing the sky regions covered by LHAASO (cyan shaded area) and by SWGO, assuming an observatory latitude close to $23^{\rm o}\,$S (orange shaded area). The map is plotted in Galactic coordinates. \label{figFoVs}}
\end{figure}
Evidence based on existing observations has firmly demonstrated that GRBs can produce energetic radiation and also that this spectral component may be associated with the elusive prompt emission (e.g. \cite{Asano09}). The observation of early high energy photons, together with hints of VHE detection of short GRBs, like in the case of GRB~160821B, challenges the external shock model predictions and suggests that additional mechanisms may be at work. Covering the VHE window, particularly in the early emission phase, will be a crucial requirement for the development of more advanced models. The ability to characterise the earliest properties of VHE emission will be fundamental both to improve the time-domain investigation of Multi-Messenger triggers, as well as to offer high quality follow-up triggers that, as demonstrated by H.E.S.S. and further boosted by the upcoming investigations with CTA, can track the VHE evolution up to several hours in the afterglow \cite{Abdalla19}.

The large FoV and the nearly continuous operating time of EAS arrays make this type of instruments an ideal facility to survey the sky looking for fast transient sources. Their ability to cover large sky areas will help constraining the VHE properties of early GRB emission, with new implications on the involved radiation mechanisms. However, in order to observe events located at Cosmological distances, they need to work effectively in the sub-TeV energy domain and, therefore, to be located in high altitude sites. The Large High Altitutde Air Shower Observatory (LHAASO) \cite{LHAASOpaper}, in the Northern hemisphere, and SWGO, from the Southern hemisphere, have the potential to carry out a VHE monitoring program that will cover a large fraction of the visible sky, as illustrated in Fig.~\ref{figFoVs}. If extended in the sub-TeV domain, this nearly constant scanning of the sky will help clarifying the VHE properties of GRBs by assessing the existence of spectral components that, though predicted in well justified models, are hard to detect with present day instruments, particularly in the prompt phase. In addition, it will offer a new window to identify sources of energetic transients, like gravitational waves and high energy neutrinos. Indeed, the identification of possible counterparts to alerts issued by continuously operating experiments, such as {\it IceCube} and {\it LIGO/VIRGO}, will undoubtedly benefit from the existence of a network of VHE monitoring programs, able to detect energetic transients and, thus, to further refine the investigation of their sources, placing constraints on their possible association and on the theoretical interpretation of their overall activity.

\section*{Acknowledgments}
This work was partly performed under project PTDC/FIS-PAR/29158/2017, Funda\c{c}\~ao para a Ci\^encia e Tecnologia. The SWGO Collaboration acknowledges the support from the agencies and organizations listed
here: \url{https:www.swgo.org/SWGOWiki/doku.php?id=acknowledgements}.

\clearpage
\section*{Full Authors List: \Coll\ Collaboration}
%
%
\scriptsize
\noindent
P.~Abreu$^1$,
A.~Albert$^2$,
E.\,O.~Ang\"uner$^3$,
C.~Arcaro$^4$,
L.\,H.~Arnaldi$^5$,
J.\,C.~Arteaga-Vel\'azquez$^6$,
P.~Assis$^1$,
A.~Bakalov\'a$^7$,
U.~Barres\,de\,Almeida$^8$,
I.~Batkovi\'c$^4$
% ...
\vspace{1cm}

\noindent
$^{1}$ {Laborat\'orio de Instrumenta\c{c}\~ao e F\'{i}sica Experimental de Part\'{i}culas (LIP), Av. Prof. Gama Pinto 2, 1649-003 Lisboa, Portugal\\}
$^{2}$ {Physics Division, Los Alamos National Laboratory, P.O. Box 1663, Los Alamos, NM 87545, United States\\}
$^{3}$ {Aix Marseille Univ, CNRS/IN2P3, CPPM, 163 avenue de Luminy - Case 902, 13288 Marseille cedex 09, France\\}
$^{4}$ {University of Padova, Department of Physics and Astronomy \& INFN Padova, Via Marzolo 8 - 35131 Padova, Italy\\}
$^{5}$ {Centro At\'omico Bariloche, Comisi\'on Nacional de Energ\'{i}a At\'omica, S. C. de Bariloche (8400), RN, Argentina\\}
$^{6}$ {Universidad Michoacana de San Nicol\'as de Hidalgo, Calle de Santiago Tapia 403, Centro, 58000 Morelia, Mich., M\'exico\\}
$^{7}$ {FZU, Institute of Physics of the Czech Academy of Sciences, Na Slovance 1999/2, 182 00 Praha 8, Czech Republic \\}
$^{8}$ {Centro Brasileiro de Pesquisas F\'{i}sicas, R. Dr. Xavier Sigaud, 150 - Rio de Janeiro - RJ, 22290-180, Brazil\\}
$^{9}$ {Academic Department of Physics – Faculty of Sciences – Universidad Nacional de San Antonio Abad del Cusco (UNSAAC), Av. de la Cultura, 733, Pabell\'on C-358, Cusco, Peru\\}
$^{10}$ {Instituto de F\'{i}sica, Universidad Nacional Aut\'onoma de M\'exico, Sendero Bicipuma, C.U., Coyoac\'an, 04510 Ciudad de M\'exico, CDMX, M\'exico \\}
$^{11}$ {Dipartimento di Fisica, Universit\`a degli Studi di Torino, Via Pietro Giuria 1, 10125, Torino, Italy\\}
$^{12}$ {RWTH Aachen University, Physics Institute 3, Otto-Blumenthal-Stra{\ss}e, 52074 Aachen, Germany \\}
$^{13}$ {University of Maryland, College Park, MD 20742, United States\\}
$^{14}$ {Durham University, Stockton Road, Durham, DH1 3LE, United Kingdom\\}
$^{15}$ {Astronomical Observatory, University of Warsaw, Aleje Ujazdowskie 4, 00478 Warsaw, Poland\\}
$^{16}$ {Facultad de Ciencias en F\'{i}sica y Matem\'aticas UNACH, Boulevard Belisario Dom\'{i}nguez, Km. 1081, Sin N\'umero, Ter\'an, Tuxtla Guti\'errez, Chiapas, M\'exico\\}
$^{17}$ {Facultad de Ciencias Exactas, Universidad Nacional de Salta, Avda. Bolivia N$^\mathrm{o}$ 5150, (4400) Salta Capital, Argentina\\}
$^{18}$ {Department of Aerospace Science and Technology, Politecnico di Milano, Via Privata Giuseppe La Masa, 34, 20156 Milano MI, Italy\\}
$^{19}$ {Facultad de Ciencias F\'{i}sico Matem\'aticas, Benem\'erita Universidad Aut\'onoma de Puebla, C.P. 72592, M\'exico\\}
$^{20}$ {Instituto de Astronomia y Fisica del Espacio (IAFE, CONICET-UBA), Casilla de Correo 67 - Suc. 28 (C1428ZAA), Ciudad Aut\'onoma de Buenos Aires, Argentina\\}
$^{21}$ {Universidad de Guadalajara, Blvd. Gral. Marcelino Garc\'{i}a Barrag\'an 1421, Ol\'{i}mpica, 44430 Guadalajara, Jal., M\'exico\\}
$^{22}$ {Instituto de F\'{i}sica de S\~ao Carlos, Universidade de S\~ao Paulo, Avenida Trabalhador S\~ao-carlense, n$^\mathrm{o}$ 400, Parque Arnold Schimidt - CEP 13566-590, S\~ao Carlos - S\~ao Paulo - Brasil\\}
$^{23}$ {INFN - Roma Tor Vergata and INAF-IAPS, Via del Fosso del Cavaliere, 100, 00133 Roma RM, Italy\\}
$^{24}$ {Dept. of Physics and CCTVal, Universidad Tecnica Federico Santa Maria, Avenida España 1680, Valpara\'{i}so, Chile\\}
$^{25}$ {Universit\"at W\"urzburg, Institut f\"ur Theoretische Physik und Astrophysik, Emil-Fischer-Str. 31, 97074 W\"urzburg, Germany\\}
$^{26}$ {Department of Physics, and the Wisconsin IceCube Particle Astrophysics Center (WIPAC), University of Wisconsin, 222 West Washington Ave., Suite 500, Madison, WI 53703, United States\\}
$^{27}$ {TUBITAK Space Technologies Research Institute, ODTU Campus, 06800, Ankara, Turkey\\}
$^{28}$ {Friedrich-Alexander-Universit\"at Erlangen-N\"urnberg, Erlangen Centre for Astroparticle Physics, Erwin-Rommel-Str. 1, D 91058 Erlangen, Germany\\}
$^{29}$ {Tecnologico de Monterrey, Escuela de Ingenier\'{i}a y Ciencias, Ave. Eugenio Garza Sada 2501, Monterrey, N.L., 64849, M\'exico\\}
$^{30}$ {Max-Planck-Institut f\"ur Kernphysik, P.O. Box 103980, D 69029 Heidelberg, Germany\\}
$^{31}$ {Universit\`a di Napoli “Federico II”, Dipartimento di Fisica “Ettore Pancini”,  and INFN Napoli, Complesso Universitario di Monte Sant'Angelo - Via Cinthia, 21 - 80126 - Napoli, Italy \\}
$^{32}$ {University of Granada, Campus Universitario de Cartuja, Calle Prof. Vicente Callao, 3, 18011 Granada, Spain\\}
$^{33}$ {IFLP, Universidad Nacional de La Plata and CONICET, Diagonal 113, Casco Urbano, B1900 La Plata, Provincia de Buenos Aires, Argentina\\}
$^{34}$ {University of Utah, 201 Presidents' Cir, Salt Lake City, UT 84112, United States\\}
$^{35}$ {Michigan Technological University, 1400 Townsend Drive, Houghton, MI 49931, United States\\}
$^{36}$ {Dipartimento di Fisica e Astronomia "E. Majorana", Catania University and INFN, Catania, Italy\\}
$^{37}$ {APC--IN2P3/CNRS, Universit\'e de Paris, Bâtiment Condorcet, 10 rue A.Domon et L\'eonie Duquet, 75205 PARIS CEDEX 13, France\\}
$^{38}$ {University of Leicester, University Road, Leicester LE1 7RH, United Kingdom\\}
$^{39}$ {Department of Physics, University of Trieste and INFN Trieste, via Valerio 2, I-34127, Trieste, Italy \\}
$^{40}${Centro de Investigaci\'on en Computaci\'on, Instituto Polit\'ecnico Nacional, Av. Juan de Dios B\'atiz S/N, Nueva Industrial Vallejo, Gustavo A. Madero, 07738 Ciudad de M\'exico, CDMX, M\'exico\\}
$^{41}$ {Department of Physics and Mathematics, Universidad de Monterrey, Av. Morones Prieto 4500, San Pedro Garza Garc\'{i}a 66238, N.L., M\'exico\\}
$^{42}$ {Department of Physics and Astronomy, University of Alabama, Gallalee Hall, Tuscaloosa, AL 35401, United States\\}
$^{43}$ {Instituto de Tecnolog\'{i}as en Detecci\'on y Astropart\'{i}culas (CNEA-CONICET-UNSAM), Av. Gral Paz 1499 - San Mart\'{i}n - Pcia. de Buenos Aires, Argentina\\}
$^{44}$ {Department of Physics, ETH Zurich, CH-8093 Zurich, Switzerland\\}
$^{45}$ {Instituto de Ciencias Nucleares, Universidad Nacional Aut\'onoma de M\'exico (ICN-UNAM), Cto. Exterior S/N, C.U., Coyoac\'an, 04510 Ciudad de M\'exico, CDMX, M\'exico\\}
$^{46}$ {Departamento de F\'{i}sica, Facultad de Ciencias B\'asicas, Universidad Metropolitana de Ciencias de la Educaci\'on, Av. Jos\'e Pedro Alessandri 774, Ñuñoa, Santiago, Chile\\}
$^{47}$ {Department of Physics, University of Seoul, 163 Seoulsiripdaero, Dongdaemun-gu, Seoul 02504, Republic of Korea \\}
$^{48}$ {Institut de recherche sur les lois fondamentales de l'Univers (IRFU), CEA, Universit\'e Paris-Saclay, F-91191 Gif-sur-Yvette, France\\}
$^{49}$ {Laboratoire Univers et Particules de Montpellier, CNRS,  Universit\'e de Montpelleir, F-34090 Montpellier, France\\}
$^{50}$ {Instituto Nacional de Astrof\'{i}sica, \'optica y Electr\'onica (INAOE), Luis Enrique Erro 1, Puebla, M\'exico\\}

%

\end{document}